\documentclass[10pt,onecolumn,letter]{IEEEtran}

  \usepackage[pdftex]{graphicx}

\usepackage[cmintegrals]{newtxmath}

\usepackage{amsmath}
\usepackage{multirow}
\usepackage{color}
\usepackage{scalerel}
\usepackage[dvipsnames]{xcolor}
\usepackage{tikz}
\usetikzlibrary{tikzmark}
\usetikzlibrary{positioning}

\def\bF{{\mathbb{F}}}

\def\Tr{\mathbb{\rm Tr}}
\newcommand\im{\mathop{\rm Im}}

\def\SD{\mathop{\rm SD}}
\def\COMP{\mathop{\rm COMP}}

\def\onoff{\mathop{\rm on-off}}

\newcommand{\sK}{\mathsf{K}}

\newtheorem{theorem}{Theorem}

\newtheorem{lemma}{Lemma}

\def\Label{\label}

\begin{document}
\title{Computation-aided classical-quantum multiple access to boost network communication speeds}
\author{%
  \IEEEauthorblockN{
    Masahito Hayashi\IEEEauthorrefmark{1}\IEEEauthorrefmark{2}
    and
    \'{A}ngeles V\'{a}zquez-Castro\IEEEauthorrefmark{3}}\\
  \IEEEauthorblockA{\IEEEauthorrefmark{1}%
    $\{$Shenzhen Institute for Quantum Science and Engineering, \\
Guangdong Provincial Key Laboratory of Quantum Science and Engineering$\}$ \\
    Southern University of Science and Technology, Shenzhen 518055, 
    China}\\
  \IEEEauthorblockA{\IEEEauthorrefmark{2}%
    Graduate School of Mathematics, Nagoya University, Nagoya, 464-8602, Japan}\\
\IEEEauthorblockA{\IEEEauthorrefmark{3}%
   $\{$the Department of Telecommunications and Systems Engineering, \\
The Centre for Space Research (CERES) of Institut d'Estudis Espacials de Catalunya (IEEC-UAB)$\}$ \\
    Autonomous University of Barcelona,
Barcelona, Spain}\\
e-mail: {hayashi@sustech.edu.cn, angeles.vazquez@uab.es}
}

\maketitle

\begin{abstract}
A multiple access channel (MAC) consists of multiple senders simultaneously transmitting their messages to a single receiver. For the classical-quantum case (cq-MAC), achievable rates are known assuming that all the messages are decoded, a common assumption in quantum network design. However, such a conventional design approach ignores the global network structure, i.e., the network topology. 
When a cq-MAC is given as a part of quantum network communication, this work shows that computation properties can be used to boost communication speeds with code design dependently on the network topology. We quantify achievable quantum communication rates of codes with computation property for a two-sender cq-MAC. When the two-sender cq-MAC is a boson coherent channel with binary discrete modulation, we show that it achieves the maximum possible communication rate (the single- user capacity), which cannot be achieved with conventional design. Further, such a rate can be achieved by different detection methods: quantum (with and without quantum memory), on-off photon counting and homodyne (each at different photon power). Finally, we describe two practical applications, one of which cryptographic.
 \end{abstract}


\section{Introduction}
Recently, quantum communication has been actively studied \cite{Hua-Ying2021,Pirandola,Azuma,Chen2021}.
For practical use of quantum communication among many users, 
we need to establish large quantum networks involving many number of nodes
\cite{Dynes,Elliott,Joshieaba0959,Kimble,Reiserer,Wehner,Satoh,Devitt}. 
However, many of their theoretical studies
do not reflect the network topology.
To see this, as a typical part of a network, we focus on a two-user multiple access channel (MAC). 
It consists of two spatially separated senders aiming to transmit simultaneously their information messages to a single receiver. 
It appears in the uplink from many terrestrial terminals to an aerial or satellite node. 
Hence, designing a code for MAC, i.e., a MAC code, is essential for building a network.
Existing studies investigated MAC codes for classical-quantum MAC, (cq-MAC),
and clarified the capacity region \cite{Winter}.
That is, these studies maximized the transmission rates under the conventional assumption that
all messages from multiple senders are decoded.
We call 
codes under such conventional assumption simultaneously-decodable (SD) codes.

In general, the network topology may include bottlenecks, 
in which case, the information flow requires to be 
network-coded to achieve the network capacity as shown by Ahlswede, Cai, Li and Yeung in \cite{Ahlswede2000}. 
Since a cq-MAC is composed of a quantum signal
from two senders, which arrive to only one receiver, 
as shown later, it becomes a bottleneck of a quantum network dependently on its topology.
In this case, to boost the communication rates over the network,
we need to resolve the bottleneck, which require a network code that takes into account the network topology (which SD codes do not consider). 
However, while several studies consider quantum network coding for noiseless channels under several topological assumptions \cite{Hayashi2007,PhysRevA.76.040301,Kobayashi2009,Leung2010,Kobayashi2010,
Kobayashi2011,JFM11,OKH17,OKH17-2,Lu,SH18-2}, 
no existing results apply the network coding concept to the cq-MAC 
composed of imperfect channels including lossy channels
dependently on the network topology.

\begin{figure}[tbh]
    \centering
    \includegraphics[scale=0.4]{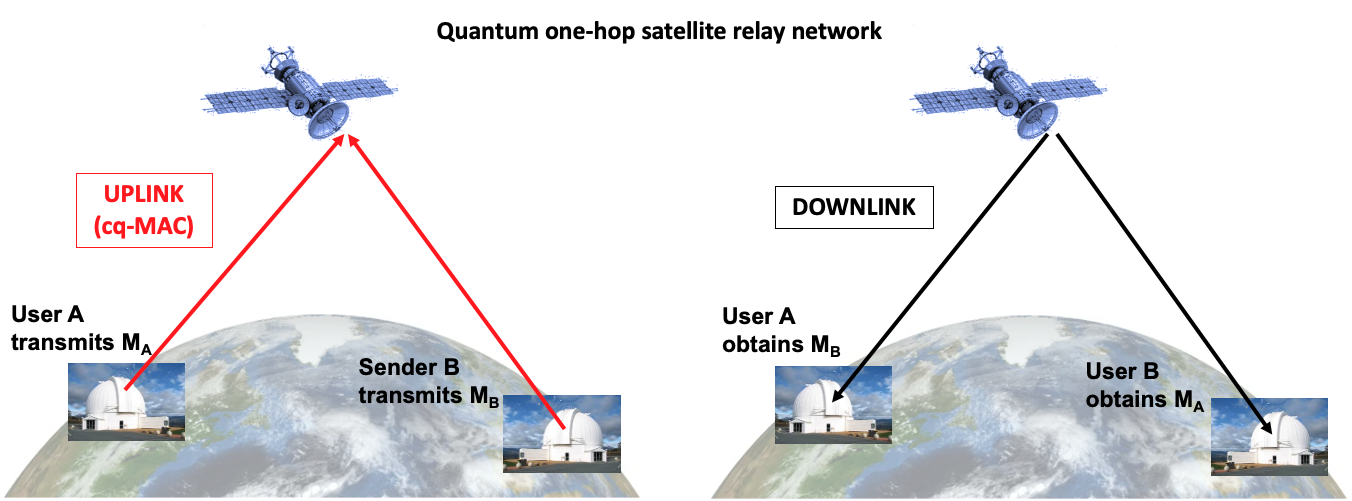}
    \caption{Physical example of quantum one-hop relay network model. 
    In this example, the relay is a satellite. The two senders A and B are located in distant places so that they cannot directly communicate to each other. In Phase 1, the two senders access simultaneously the satellite. 
    Hence, the uplink is a cq-MAC (the network bottleneck) that has two senders A and B and a single decoding system on the satellite. In Phase 2, the relay communicates to each sender via two point-to-point cq-channels towards senders A and B. In this work we show that conventional decoding is not optimal (i.e. cannot achieve the capacity of the network) and introduce computation (COMP) codes that allow to decode the actual sum of (quantized) electromagnetic fields in the cq-MAC of Phase 1, enabling network capacity to be achieved.}
    \label{OneHopNetworkScenario}
\end{figure}

Here, 
to see how conventional codes underperform (i.e. do not achieve capacity) due to the bottleneck of the cq-MAC on a concrete network,
we focus on the quantum one-hop relay network model, which is a simple,
yet relevant network topology, and is composed of a relay node (e.g. aerial vehicles or satellites) in addition to
two senders A and B (see Fig. \ref{OneHopNetworkScenario}).
The two senders A and B are located in distant places so that they cannot directly communicate to each other.
They can access only to the relay node (illustrated as a satellite in Fig. \ref{OneHopNetworkScenario}).
Thus, this network model has two types of channels.
One is the uplink, a cq-MAC (the network bottleneck) that has two senders A and B and one receiver, the relay node.
The other is the downlink that is composed of two point-to-point cq-channels from the relay node 
to each of A and B.
The aim of this network is that the two senders A and B exchange their messages $M_A$ and $M_B$ via the relay. 
Although the use of cq-MAC saves time, 
the quantum signal transmitted from one sender behaves as a noise for 
the quantum signal from another sender, which is the weak point of cq-MAC.
Since the point-to-point cq-channels in the downlink have no such weak point,
only the uplink is 
the bottleneck of the quantum one-hop relay network model.
Note that, physically, when we employ the bosonic 
quantized electromagnetic fields, 
the cq-MAC in the uplink is given as 
interference of electromagnetic fields.

\begin{figure}[tbh]
    \centering
    \includegraphics[scale=0.39]{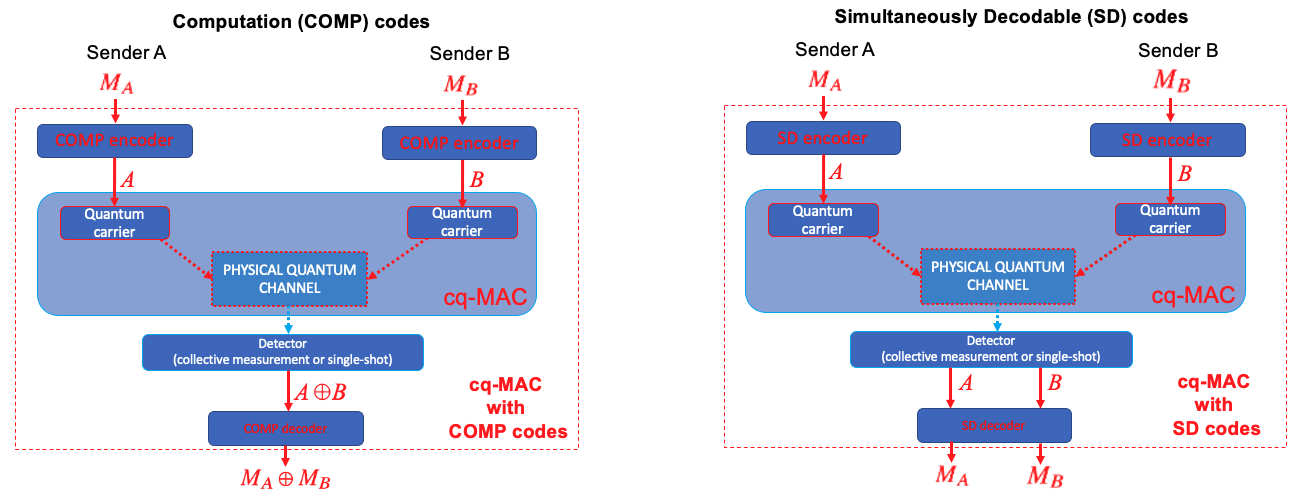}
    \caption{Illustration of the operation of COMP codes proposed in this work, compared to conventional SD codes when two senders simultaneously transmitting their messages over a cq-MAC. The diagram on the left illustrates the concept of COMP codes.    The main objective of our work is the design of the required COMP codes when only the modulo sum $M_A \oplus M_B$ needs to be reliably transmitted over the cq-MAC. For comparison, the conventional approach (\cite{Winter}\cite{Yen}) is shown on the right: the receiving node decodes both messages, $M_A$ and $M_B$ using conventional 
    simultaneous-decodable codes, denoted here as SD codes.}
    \label{COMPconcept}
\end{figure}

In the conventional relying method, 
the receiver at the relay decodes both messages in the unplink, 
and sends $M_A$ to Sender B and message $M_B$ to Sender A in the downlink.
This relying method employs a SD code in the uplink, and is called direct forward (DF) because the relay directly decodes-and-forwards the messages. 
However, 
the relay can operate in a different way. 
To show a different relaying method 
to resolve the topological bottleneck,
we need to introduce computation (COMP) codes over the cq-MAC.
The operation of COMP codes compared to SD codes in a cq-MAC is illustrated in Fig. \ref{COMPconcept}.

On the left, the use of COMP codes is shown 
when two senders transmit their messages $M_A$ and  $M_B$ over a cq-MAC but only the modulo sum $M_A \oplus M_B$ is required to be reliably decoded. 
For comparison, the conventional SD codes in \cite{Winter,Yen,Wilde} is shown on the right, 
where the two messages $M_A$ and  $M_B$ would be decoded simultaneously.

We observe that in principle, either SD or COMP codes can be used in our quantum 
one-hop network model. In other words, the satellite in Fig. 1 can be designed to use either SD or COMP codes.
However, since the aim of the network is the exchange of messages between nodes A and B,
it is sufficient that the receiver at the relay node decodes only the modulo sum $M_A \oplus M_B$
in the uplink,
where both messages are regarded as vectors on a finite field $\bF_p$.
This is because 
both senders decode the other's message by using the modulo operation
after the relay broadcasts the recovered modulo sum $M_A \oplus M_B$ to both senders
in the downlink.
For example, Sender A can decode $M_B$ by $(M_A \oplus M_B) \ominus M_A$.
In this paper, we use $\oplus$ and $\ominus$
to express the modulo plus and minus in the finite field $\bF_p$
to distinguish it from the plus and minus in $ \mathbb{C}$, respectively.
This method employs a COMP code in the uplink, and is called 
computation-and-forward (CAF) strategy because the relay forward the result of the computation of the messages. 
In the satellite network model in Fig. \ref{COMPconcept}, our cq-MAC is given over quantized electromagnetic fields, a typical example of a bosonic system, and
this method boosts the capacity by harnessing photon energy from the natural computation properties existing in 
the coherent interference of quantized electromagnetic fields. 

In the classical case, our proposed type of codes are said to provide reliable physical layer network coding
because the code is directly applied to the physical signal naturally combined in the channel.
However, while
our proposed strategy seemingly corresponds to the quantum analogue of physical layer network coding (which has been extensively studied
\cite{Liew06,Nazer06,Popovski06,Nazer2007,Nazer08,Nazer11,NazerRel11,Nazer16,Ullah,Takabe,Takabe2,HV}), in fact it does not. 
The reason is that in traditional classical networks and Internet, physical networks and their logical abstractions are purposely separately designed and managed (e.g. the former by engineers the latter by computer scientists). 
Fully quantum networks, however, no longer have such separation because its developing stage is still in pioneer days
before establishing the division 
between physical and logical parts.
To extract the performance of the quantum network,
we need to seamlessly design the communication method over the quantum network.

%
%

In this work, we study the codes for the cq-MAC that take into account the network topology, i.e. COMP codes for the cq-MAC, and obtain a lower bound of achievable rates. We then apply them over cq-MAC with boson coherent states \cite{Yen,Wilde}
and quantify the gains in reliable rates with respect to conventional SD codes 
for several 
detection strategies. 
We show that COMP codes boost communication rates by taking advantage of the computation properties inherent in the coherent quantum interference (of quantized electromagnetic fields).
When our method is applied to quantized electromagnetic fields,
COMP codes are interpreted to gather photon energy from interference instead of treating it as noise (like conventional codes do).
In terms of relying strategy for our network of reference (Fig. \ref{OneHopNetworkScenario}), our results show that the CAF relaying strategy is preferred over the conventional DF strategy.
In addition, although the merit of COMP codes depends on the network topology,
there are many other types of topologies whose bottlenecks can be resolved by our method,
such as the well known butterfly network \cite{Ahlswede2000,Hayashi2007,PhysRevA.76.040301,OKH17}, 
which is described in Appendix A.

To show the importance of the seamless design, 
in addition to the CAF strategy, 
we introduce another significant application which shows the wide applicability of our proposed COMP codes. 
For this aim, we abstract away a quantum physical server as a logical server node. We then show that
COMP codes can be applied to symmetric private information retrieval with such
servers.
 Private Information Retrieval (PIR) with multiple servers is a method for a user to download a
file from non-communicating servers each of which contains
the copy of a classical file set while the identity of the
downloaded file is not leaked to each server. 
Symmetric PIR (SPIR) with multiple servers 
achieves the above task without leakage of 
other files to the user. Hence, this is another contribution of our work as a similar result is not yet existing in the literature, which has only treated noiseless channels \cite{SJ} \cite{SH}.

\section{Results}
\subsection{Achievable rates of COMP codes}



For simplicity, we assume the two senders of the cq-MAC, denoted as $\cal W$, having the same codebook with the alphabets ${\cal A}$ and ${\cal B}$ defined in $\bF_p$. 
Each message $M_k^n$, $k=A, B$ is assumed to be chosen independently and uniformly from $\mathcal{M}_k^n$. We define a $(2^{nR}, n)$ COMP code as consisting of two message sets that are equivalent to $\mathcal{M}^n=\bF_p^\ell$ with $|\bF_p^\ell|\cong [2^{n R}]$. 
Two encoders $\phi^n_1$ and $\phi^n_2$ map each message $M_k^n \in  \mathcal{M}^n$ to a sequence $\phi^n_k(M^n_k) \in \mathbb{C}^n$ and $k=1, 2$. Then, after measuring the received quantum system, the receiver estimates a linear combination of $M^n_1\oplus M^n_2 \in  \mathcal{M}^n$.

The rate $R$ is achievable if there exists a sequence of 
$(2^{nR}, n)$ computation codes such that $\lim_{n\rightarrow \infty} P^n_{\mbox{\tiny COMP}}= 0$, with $P^n_{\mbox{\tiny COMP}}$ the error probability. 
The supremum of achievable rates for computation is called the computation rate and is denoted by ${C}_{\COMP}$. 
To get its lower bound instead of ${C}_{\COMP}({\cal W})$,
we denote by $I(A\oplus B; Y)_{P_U \times P_U}$
the mutual information 
when $A$ and $B$ are subject to the uniform distribution $P_U$ independently, i.e.,
$P_A \times P_B=P_U \times P_U$.
The capacity ${C}_{\COMP}({\cal W})$ is evaluated in
the following theorem. 
\begin{theorem}\label{T1}
we have the following lower and upper bounds of ${C}_{\COMP}({\cal W})$.
\begin{align}
\underline{{C}}_{\COMP}({\cal W}) &:=I(A\oplus B; Y)_{P_U \times P_U} \le {C}_{\COMP}({\cal W})\label{Eq1}\\
\overline{{C}}_{\COMP}({\cal W}) &:=
\min \Big(\max_{P_A , b \in \bF_p}I(A; Y|B=b)_{P_A} , 
\max_{P_B, a\in \bF_p}I(B; Y|A=a)_{P_B} \Big)
\ge {C}_{\COMP}({\cal W}),\label{Eq2}
\end{align}
where 
the distribution of $A$ and $B$ are limited to the uniform distribution in \eqref{Eq1}
and
the maximum in \eqref{Eq2} is taken among distributions 
$P_A $ or $P_B$ of $A$ or $B$.
\end{theorem}
Theorem \ref{T1} is 
proved in Appendix C by combining 
the method for degraded channel \cite{Ullah},
the operator inequality developed in \cite{HN},
the technique based on Toeplitz matrix \cite{TH13}, and 
affine codes.
Note that a conventional SD code is defined as a $(2^{nR_A}, 2^{nR_B}, n)$ with two message sets $[2^{nR_k}]$, $k=A, B$ and two encoders $\phi^n_1$ and $\phi^n_2$ which use \emph{different codes} to map each message  $M_k^n \in  [2^{nR_k}]$ to a sequence $\phi^n_k(M^n_k) \in \mathbb{C}^n$, $k=A, B$. Each message $M_k^n$, $k=A, B$ is assumed to be chosen independently from $[2^{nR_k}]$. 
In this case, the receiver outputs two estimations, of $M^n_{A}$ and of $M^n_{B}$. 
The pair $(R_A, R_B)$ is achievable for simultaneous-decodable code if there exists a sequence of $(2^{nR_A}, 2^{nR_B}, n)$ simultaneous-decodable codes such that $\lim_{n\rightarrow \infty} P^n_{\mbox{\tiny SD}} = 0$, 
where $P^n_{\mbox{\tiny SD}}$ is the error probability. 
In this case, the rates are described by the capacity region for simultaneous-decodable code, ${\cal C}({\cal W})$,
\cite{Yen, Winter}
and interference is treated as noise.  
When a SD code is used, the rate is limited as
\begin{align}
C_{\SD}({\cal W})&:=
\max_{(R_A,R_B) \in {\cal C}({\cal W})}\min(R_A,R_B)
=
\max_{ P_A \times P_B}
\min \Big(
I(A;Y|B)_{P_A \times P_B},
I(B;Y|A)_{P_A \times P_B},
\frac{1}{2}I(AB;Y)_{P_A \times P_B}
\Big).
\label{mincutSD}
\end{align}

In fact, to achieve the rates $\underline{{C}}_{\COMP}({\cal W})$ and $C_{\SD}({\cal W})$,
the decoder needs to use collective measurement across $n$ receiving signal states.
This measurement requires the technologies to store many receiving quantum signal states
and perform a quantum measurement across many quantum signal states. 
To avoid use of quantum memory, the receiver needs to measure the receiving states individually.

As a simple case, we consider the 
situation where the receiver applies the same measurement $\Pi$ to all receiving signals.
We denote the measurement outcome by $Z$.
Then, we can define the mutual information $I(A\oplus B; Z)_{\Pi}$.
It is natural that the receiver's decoding operation is limited to 
the classical data processing over the collection of the outcomes of the same measurement $\Pi$.
This method does not require quantum memory.
In this case, we have the following achievable rate by using the result of classical COMP codes 
$$\underline{{C}}_{\COMP,\Pi}({\cal W}) :=I(A\oplus B; Z)_{\Pi,P_U \times P_U}.$$
Optimizing the choice of the measurement, we obtain the rate 
$\underline{{C}}_{\COMP}^c({\cal W}):=\max_\Pi \underline{{C}}_{\COMP,\Pi}({\cal W})$
achieved without quantum memory.
Besides, the inequality $\underline{{C}}_{\COMP}^c({\cal W})
\le \underline{{C}}_{\COMP}({\cal W})$ holds,
the equality holds if and only if the density matrices
$\{\frac{1}{p}\sum_{a \in \bF_p}  W_{a,b\ominus a}\}_{b \in \bF_p} $
are commutative with each other.
Hence, in this case the receiver does not need to use collective measurement across $n$ receiving signal states.
When a SD code is used and the receiver applies the measurements $\Pi$ to all receiving signal states,
the rate is limited as
\begin{align}
C_{\SD, \Pi}({\cal W})&:=
\max_{ P_A \times P_B}
\min \Big(
I(A;Z|B)_{\Pi,P_A \times P_B},
I(B;Z|A)_{\Pi,P_A \times P_B},
\frac{1}{2}I(AB;Z)_{\Pi,P_A \times P_B}
\Big).
\label{mincutSD_M}
\end{align}
Optimizing the choice of the measurement, we obtain the rate 
${{C}}_{\SD}^c({\cal W}):=\max_\Pi C_{\SD, \Pi}({\cal W})$.

\subsection{Application to BPSK discrete modulation over lossy bosonic channel}
As a practical application of our results, we now study the cq-MAC over coherent states with binary phase shift keying (BPSK) discrete modulation. In this case, physically,
COMP codes match computation properties of (quantized) physical electromagnetic carriers. 
In Methods, we present the detailed description of the physical models of the coherent cq-MAC. 
These allow us to derive analytically the rate with 
collective quantum detection
and also the rates with photon counting (on-off) and homodyne detections. 
Since the collective quantum detection needs to be applied across multiple receiving quantum signal systems,
it requires quantum memory.
On the other hand, photon counting (on-off) and homodyne detections
can be applied to single receiving quantum signal system
so that they do not need quantum memory.


\begin{figure}[tbh]
    \centering
    \includegraphics[scale=0.4]{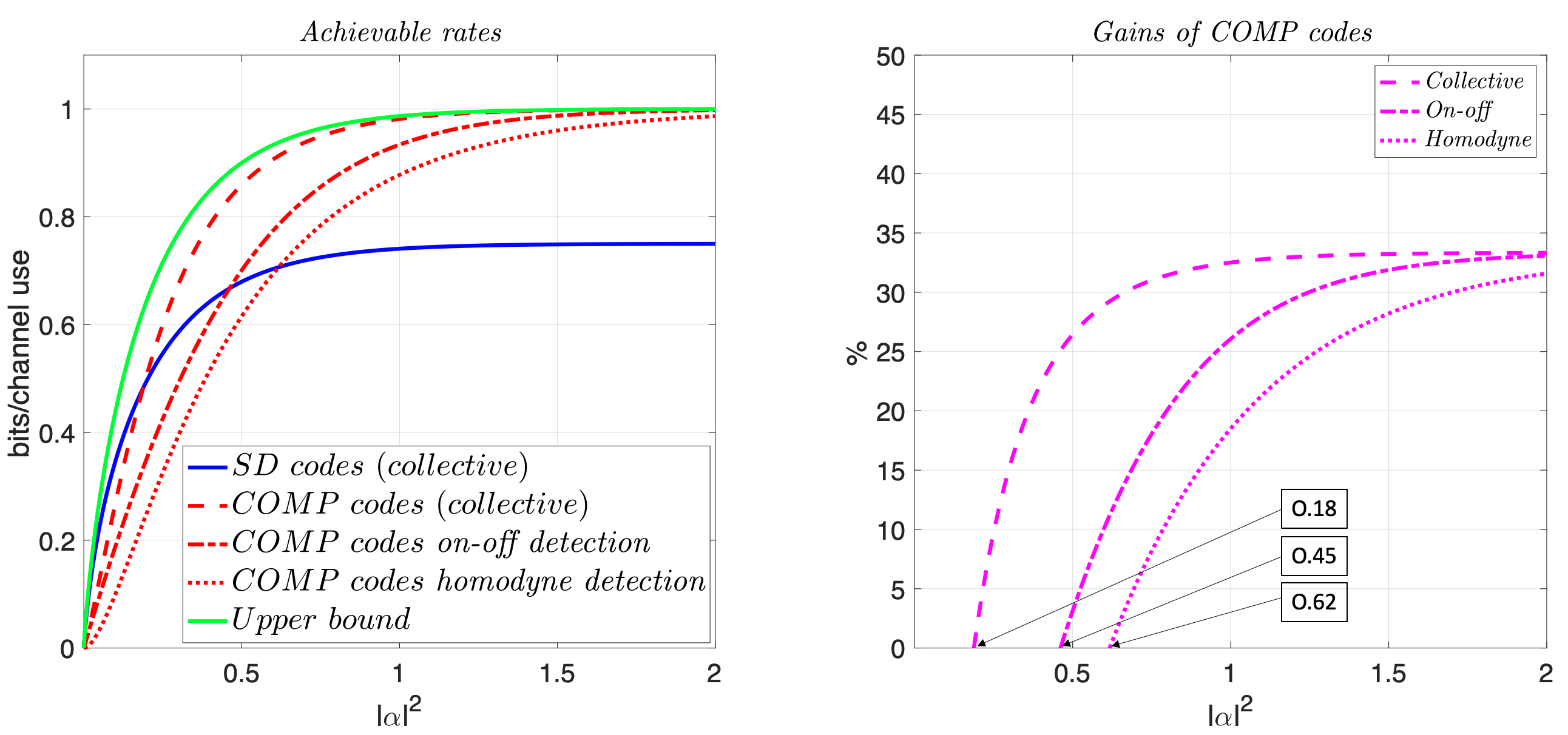}
    \caption{Left: Achievable communication rates with collective measurement, with SD codes, expression  \eqref{Expression_Achievable_rates_SD} (in blue) and with COMP codes for different measurements: collective as expression \eqref{eq:rate_COMP} (in red), on-off as expression \eqref{eq:comprateBPSK} (in dashed red) and homodyne as expression \eqref{eq:Expression_Achievable_rates_COMP_hom} (in dashed point red). Right: the corresponding gains of COMP codes with different measurements over SD codes (up to $33.33\%$).}
    \label{InterferencePlots}
\end{figure}

Achievable rates in Fig. \ref{InterferencePlots} (left) demonstrate the superiority of COMP codes for the bosonic cq-MAC network, as they achieve the 
upper bound of the capacity (1 bit per channel use for BPSK), 
while SD codes do not. 
Specifically, the rate of conventional codes with collective measurement, $C_{SD}(W_\alpha),$ (which is maximum for uniform priors) achieves $I_{\max,\infty}/2 = (\eta(0.5) + \eta(0.25) + \eta(0.25))/2=0.75$ bits per channel use at $|\alpha|^2  \approx 1$, where
$\eta(x)=- x\log_2 x$.
However, COMP codes with collective measurement achieve 
the transmission rate
$\log_2(2) = 1$ bits per channel use at $|\alpha|^2  \approx 1$. 
Hence, COMP codes outperform conventional SD codes with gains up to $33.33\%$ as shown in Fig. \ref{InterferencePlots} (right), where we have defined gain between two rates, $R_x$ and $R_y$ as $100\frac{R_x-R_y}{R_y}$.
In addition, the on-off measurement also achieves 
the transmission rate $1$ with COMP codes but at a slower speed than with collective measurement, 
and it happens also at $|\alpha|^2 \approx 1.5$. 
Even the conventional homodyne detection achieves 
the transmission rate $1$ with COMP codes at $|\alpha|^2 \approx 2$.
The gains of our method for COMP codes over SD codes with collective measurement
are positive only after a threshold photon power for any type of measurement. 
The thresholds of the collective measurement, the on-off measurement
and the homodyne detection are $|\alpha|^2 =0.18,0.45$, and $0.62$, respectively.
The intuition of this result is that the quantum nature of the received signal
offers an advantage for the decoding of the sum of the physical signal over 
SD codes
above such threshold photon power while it does not below the threshold.

As a useful reference, the single-sender Holevo quantity gives the upper bound \eqref{Eq2}
of the capacity of COMP code, 
as shown in Fig. \ref{InterferencePlots} (left, green line). 
We observe that COMP codes with collective measurement
is very close to the upper bound from a realistic photon average photon number of 1.


\subsection{Applications to quantum networking tasks}
Finally we show the practical relevance of our results describing 
the realistic application of COMP codes for two concrete quantum networking tasks: 
information exchange over the quantum one-hop relay network model
and private information retrieval (SPIR protocol) between one user and two servers. 
While the former is the quantum analog of a classically known strategy as computation-and-forward, 
the latter is a novel application and therefore we provide a more detailed description.

\subsubsection{Information Exchange between Alice and Bob}
\begin{figure}[tbh]
    \centering
    \includegraphics[scale=0.3]{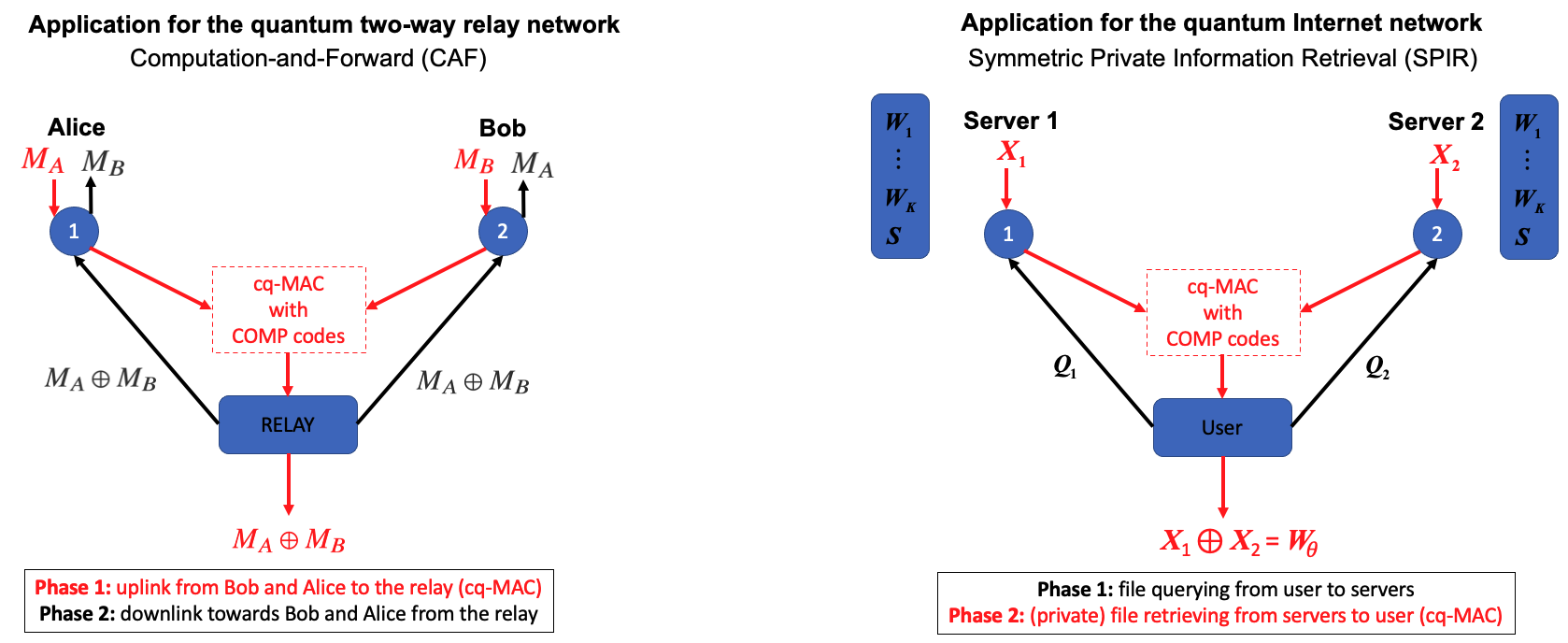}
    \caption{Left: schematic illustration of the quantum compute-and-forward relaying strategy, whereby Bob and Alice exchange their messages over a cq-MAC channel with COMP codes. Right: schematic illustration of our method for symmetric private information retrieval (SPIR) with two servers based on cq-MAC with COMP codes. Observe the very different role of the cq-MAC in the two applications while they employ the same network model,
    the quantum one-hop relay network model.}
    \label{Applications}
\end{figure}

Here, to clarify the difference from the application to SPIR, we summarize the CAF strategy.
As explained in introduction, the CAF strategy assumes the quantum one-hop relay network model, composed of a relay node (e.g. aerial vehicles or satellites) in addition to two senders, Bob and Alice, communicating over a wireless channel. 
The aim of this network is that Bob and Alice exchange their respective messages 
via the relay. 
Hence, it is sufficient that the receiver at the relay node decodes only the modulo sum $M_A \oplus M_B$
in the uplink,
where both messages are regarded as vectors on a finite field $\bF_p$.
This is because 
both senders decode the other's message by using the modulo operation
after the relay broadcasts the recovered modulo sum $M_A \oplus M_B$ to both senders by using two 
point-to-point channels.
For example, Sender A can decode $M_B$ by $(M_A \oplus M_B) \ominus M_A$. This application is schematically illustrated in Fig. \ref{Applications} (left). 
We observe that in-network computation (instead of simultaneous decoding) with COMP codes is enough to accomplish the networking task of information exchange between Bob and Alice.

\subsubsection{Private Information Retrieval}
Private Information Retrieval (PIR) with multiple servers is a method for a user
to download a
file from non-communicating servers each of which contains
the copy of a classical file set while the identity of the
downloaded file is not leaked to each server. 
This problem is trivially solved by requesting all files to one of the servers,
but this method is inefficient. 
Also, this method allows the user to get information for other files that are not intended.
Symmetric PIR (SPIR) with multiple servers is a 
protocol to achieve the above task without leakage of 
other files to the user.
If the number of servers is one, this task is the same as the oblivious transfer, which is known as a difficult task.
In a conventional setting, we assume that the number of servers and the number of files are fixed
and all files have the same size and their sizes are asymptotically large. That is,
the rate of the file size are increased linearly for the number $n$ of channel uses.
In this case, the cost of upload from the user to the servers are negligible, i.e.,
the required information rate is zero.
Only the size of download information is increased.
Therefore, we optimize the maximum rate of the file size
with respect to the number $n$ of channel uses.

When a classical noiseless channel from each server to the user is given
and shared common random numbers between servers are allowed,
the paper \cite{SJ} derived the optimal protocol and its rate. 
Also, 
when a quantum noiseless channel from each server to the user is given
and shared entangled states between servers are allowed,
the paper \cite{SH} derived the optimal protocol and its rate. 
However, no existing study address SPIR with multiple servers
when classical MAC nor cq-MAC is given as a channel from 
servers to the user.
In this paper, we focus on the case with two servers. 
That is,
the quantum one-hop relay network model is given as follows.
Two server access the cq-MAC whose receiver is the user.
The user has the respective point-to-point channels to both servers. 
This setting includes the case when the cq-MAC is a classical MAC.

As a preparation, we review the optimal method by \cite{SH} 
when noiseless classical channels from two server to the user are available.
Assume that we have $\sK$ files and the $\sK$ files messages are given  
as random variables $W_1, \ldots, W_{\sK} \in \bF_p^\ell$, which are shared by two servers.
The two severs shares uniform random variables $S \in \bF_p^\ell$.  
The optimal protocol is the following when 
the user wants to get the $\theta$-th file.
\begin{description}
\item[{\bf Upload}]\quad
The user generates $\sK$ binary random variables $(Q_{1,1}, \ldots, Q_{1,\sK}) \in \bF_2^{\sK}$
independently subject to the uniform distribution.
Then, the user sends the query $Q_i =(Q_{i,1}, \ldots, Q_{i,\sK})$ to the $i$-th server,
where $Q_{2,j}:=Q_{1,j}$ for $j \neq \theta$ and $Q_{2,\theta}:=Q_{1,\theta} \oplus 1$.
\item[{\bf Download}]\qquad
The first server sends the user $X_1=(\oplus_{j: Q_{1,j}=1}  W_j)\oplus S$.
The second server sends the user $X_2=-((\oplus_{j:  Q_{2,j}=1} W_j)\oplus S)$.
\item[{\bf Decoding}]\qquad
When $Q_{1,\theta}=1$, the user recovers the $\theta$-th file 
by $X_1 \oplus X_2= W_\theta$.
When $Q_{1,\theta}=0$, the user recovers the $\theta$-th file 
by $-(X_1\oplus X_2)= W_\theta$.
\end{description}
Combining the above idea with dense coding \cite{BW},
the recent paper \cite{SH} addressed SPIR with multiple users with noiseless quantum channel.

Under the above protocol, 
the information for $\theta$ is not leaked to each server unless the two servers communicate
each other.
Also, the user cannot obtain any information for other files 
$W_1, \ldots, W_{\theta-1}, W_{\theta+1},\ldots, W_{\sK}$.
Hence, the task for SPIR with two servers is realized by the above protocol.
In the above protocol, the amount of upload information is a constant, which is independent of
the file size. 

Now, we consider the case when 
the quantum one-hop relay network model connects two servers and one user in the above way.
Trivially, when we apply a SD code to a given cq-MAC,
we can simulate the noiseless channel from each servers.
Hence, for the download step of we can use the noiseless channel realized by 
a SD code over a given cq-MAC.
That is, in the download step, 
the $i$-th server sends $X_i$ to the user via the SD code over the given ca-MAC.
This idea can be extended to the case with COMP codes.
Since the key point of the above protocol is that the user recovers $X_1 \oplus X_2$,
we can apply our COMP code to the combination of the download step and the decoding step
in the above protocol of SPIR with two servers.
In summary, when a COMP code is employed to SPIR, 
the protocol in the download and decoding parts is changed as follows.
\begin{description}
\item[{\bf Download}]\qquad
The first server encodes $X_1=(\oplus_{j: Q_{1,j}=1}  W_j)\oplus S$, and
the second server encodes $X_2=-((\oplus_{j:  Q_{2,j}=1} W_j)\oplus S)$
by using a COMP code.
\item[{\bf Decoding}]\qquad
Using a COMP code, The user decodes $X_1 \oplus X_2$. 
When $Q_{1,\theta}=1$, the user recovers the $\theta$-th file 
by $X_1 \oplus X_2= W_\theta$.
When $Q_{1,\theta}=0$, the user recovers the $\theta$-th file 
by $-(X_1\oplus X_2)= W_\theta$.
\end{description}
Thus, when a cq-MAC is given
and shared common random numbers between servers are allowed,
the rate ${C}_{\COMP}({\cal W})$
is achievable for SPIR with two servers.
Therefore, when we employ a SD code or a COMP code to realize SPIR with two servers,
the rate of the SD code or the COMP code expresses
the rate of SPIR based on the respective method. 
We observe that in-network computation (instead of simultaneous decoding) with COMP codes is enough to accomplish the networking task of private retrieval from a user to two servers.

\section{Discussion}
The general achievement of this paper is to show that 
the maximum achievable rate (capacity) requires 
codes reflecting the network topology of a quantum network
when the quantum network has bottlenecks in the topology. 
As the main result of this paper, we have introduced 
COMP codes over the cq-MAC, which is the simplest example of bottleneck in the physical topology, 
and have derived general formulas of achievable rates.

To see the above fact in a physical concrete model, 
we have demonstrated the effect of COMP codes over 
a bosonic cq-MAC whose quantum carriers are coherent states. 
Physically, a COMP code 
gathers photon power from interference 
by using the computation properties of the (quantized) electromagnetic field,
rather than regarding interference as a noise. 
Under this model, we have derived the analytical expressions and 
computed the numerical values of achievable rates for a practical discrete binary modulation, the binary phase shift keying (BPSK). 
Our results show that COMP codes achieve the transmission rate close to $1$ 
under the above model at power $\|\alpha\|\approx 1$
while conventional codes cannot, which shows up to 33.33\% gains over 
the conventional method.
Surprisingly, not only with quantum detection, but also with suboptimal on-off and homodyne detection, 
COMP codes achieve the transmission rate close to $1$ 
under the same model, just requiring more photon power than with 
collective measurement.
Our results also show that 
the proposed method of COMP codes outperforms conventional codes only 
above a certain (small) threshold due to the quantum nature of the physical detection. 
That is, the quantum nature of the received signal offers an advantage for the decoding of the computation (sum) of the physical signal over SD codes
only above such threshold photon power while it does not below the threshold.

Finally, in order to clarify the applications, we have found that while in traditional classical networks and Internet, physical networks and their logical abstractions are purposely separately designed and managed (e.g. the former by engineers the latter by computer scientists), fully quantum networks are still in pioneer days before establishing such division 
between physical and logical parts. 
Hence, we have explained two practical applications assuming seamlessly design of communication over the quantum network.

To simply clarify the advantage of COMP codes,
this paper focuses on BPSK modulation on lossy coherent states
as a simple modulation. 
Therefore, it is an interesting remaining study
to investigate COMP codes with more practical modulations
on coherent states over noisy bosonic channels.

While we have proved that using computation properties of quantum interference can improve significantly the rates of a network, it seems reasonable to question whether such improvement could be harnessed from other quantum resources. As an example, it is well known that pre-shared entanglement can significantly boost communication rates in the regime of high thermal noise and very low photon power \cite{EA-Shor2002,EA-Holevo2002,EA-Winter2008,EA-Zhang2020,EA-Guha2020,WH}. 
This is due to the fact that entanglement-assisted communication scales with the factor $\log(1/|\alpha|^2)$ \cite{EA-Guha2020}. Hence, for the channel in our work, it is interesting to observe that 
while entanglement assisted cq-MAC is useful only for very low photon power, 
COMP codes are useful only above a photon threshold, thus making both techniques compatible and complementary.

In addition, as another future problem, we can consider the secrecy condition for 
each message against the receiver. 
That is, we impose that the receiver can recover the modulo sum, but has no information for each message.
Fortunately, the classical version of this problem has been studied in various research works \cite{Ren,He1,He2,Vatedka,Zewail,HWV}.
Therefore, we can expect to generalize the above classical result to our quantum setting.
Such a topic is another fruitful future study.

\section{Methods}
 
\subsection{Gains of COMP over boson coherent channels}
We now show the application of COMP codes over boson coherent channels of optical coherent light. It corresponds to a realistic scenario where multiple sending devices communicate with a single receiver (terrestrial, aerial or space-borne).
For a feasible implementation,
we focus on the practical discrete modulation binary phase shift keying (BPSK).
BPSK is obtained by choosing
the input complex amplitudes of Senders A and B to be 
$(-1)^a \tau^{-1/2} \alpha$ and $(-1)^b \tau^{-1/2} \alpha$, 
respectively, for $a,b \in \{0,1\}$, where the photon number constraint at each sender is given as $|\alpha|^2,$
$\tau$ is the transmittance for modes $A$ and $B$ of the channel, 
which we assume equal in both channels. More precisely, denoting the signal
modes as $\hat{a}_A$, $\hat{a}_B$ for Senders A and B, respectively and $\hat{a}_E$ the environment mode, the receiver's mode $\hat{a}_R$ is given as \cite{Yen,Wilde,EA-Guha2021}
\begin{align}
\hat{a}_R= \sqrt{\tau}\hat{a}_A+ \sqrt{\tau} \hat{a}_B+\sqrt{1-2\tau} \hat{a}_E.
\label{RecOP}
\end{align}

Then, the receiver obtains the coherent state 
\begin{align} 
W_{a,b}:=|((-1)^a + (-1)^b) \alpha \rangle \langle ((-1)^a + (-1)^b)\alpha | \label{2nd-channel}.
\end{align}
We denote this cq-MAC as ${\cal W}_\alpha.$ 
This channel can be implemented by controlling the photonic power of the input signals in both sender sides. 
Here, we explain only bounds with simple expression.
By using the constants 
$\tilde{c}_{0}:=(1+e^{-4|\alpha|^2})/2 $,
$a_{o|\alpha}:=e^{-|\alpha|^2} \sinh |\alpha|^2$, and
$a_{\overline{e}|\alpha}:=e^{-|\alpha|^2} (\cosh |\alpha|^2-1)$,
our upper and lower  bounds $\overline{{C}}_{\COMP}({\cal W}_\alpha)$ and
$\underline{{C}}_{\COMP}({\cal W}_\alpha)$
have the following simple analytical expressions;
\begin{align}
\overline{{C}}_{\COMP}({\cal W}_\alpha)&=h(a_{o|\alpha})
\label{NJ1}\\
\underline{{C}}_{\COMP}({\cal W}_\alpha)&=
H\left[
 \left(
\begin{array}{ccc}
\tilde{c}_0& \sqrt{a_{\overline{e}|2\alpha} } e^{-2|\alpha|^2}/2& 0  \\
\sqrt{a_{\overline{e}|2\alpha} } e^{-2|\alpha|^2}/2
& a_{\overline{e}|2\alpha}/2&0  \\
0& 0& a_{o| 2\alpha}/2 
\end{array}
\right)
 \right]
-\frac{1}{2} h(a_{o| 2\alpha}) .\label{NJ2}
\end{align}
Also, the achievable rate by the on-off measurement is calculated as
\begin{align}
\underline{{C}}_{\COMP,\onoff}({\cal W}_\alpha)=
h (\tilde{c}_{0})-\frac{1}{2}h(e^{-4|\alpha|^2}),\label{JO3}
\end{align}
where $h$ is the binary entropy function and $H$ is the von Neumann entropy.

\subsection{Rate $C_{\SD}({\cal W}_\alpha)$ for SD codes}
When we consider the conditional mutual information $I(A;Y|B)$,
we calculate the mutual information between the input $A$ and the output quantum system $Y$ under the condition that the variable $B$ is fixed to a certain value $b$.
Hence, it is sufficient to calculate the mutual information for the channel
$a\mapsto 
|((-1)^a + (-1)^b) \alpha \rangle \langle ((-1)^a + (-1)^b)\alpha |$.
Since $b$ is a fixed value, applying the displacement unitary 
$D(- (-1)^b\alpha)$, we have the state 
$|(-1)^a\alpha  \rangle \langle (-1)^a\alpha  |$.
That is, it is sufficient to calculate the mutual information for the channel
$a\mapsto 
|(-1)^a\alpha  \rangle \langle (-1)^a\alpha  |$.
By using two orthogonal states 
\begin{align}
|odd_{\alpha}\rangle:=(\sinh |\alpha|^2)^{-1/2}
\sum_{m=0}^{\infty} \frac{\alpha^{2m+1}}{\sqrt{(2m+1)!}}
|2m+1\rangle, \quad
|even_{\alpha}\rangle:=(\cosh |\alpha|^2)^{-1/2}
\sum_{m=0}^{\infty} \frac{\alpha^{2m}}{\sqrt{(2m)!}}
|2m\rangle ,\label{ALP}
\end{align}
and
$a_{o|\alpha}:=e^{-|\alpha|^2} \sinh |\alpha|^2$ and $a_{e|\alpha}:=
e^{-|\alpha|^2} \cosh |\alpha|^2$,
the coherent state is decomposed as
\begin{align}
|\pm\alpha\rangle
=
\sqrt{a_{e|\alpha}} |even_{\alpha}\rangle
\pm  \sqrt{a_{o|\alpha}} |odd_{\alpha}\rangle.
\end{align}
Dependently of $P_A$ and $P_B$, we have
\begin{align}
&I(A;Y|B)
=H(
P_A(0)|\alpha \rangle \langle \alpha  |
+P_A(1)|-\alpha \rangle \langle -\alpha  |) \nonumber\\
=& H\left[
 \left(
\begin{array}{cc}
a_{o| \alpha} & c(\alpha,P_A(0)) \\
c(\alpha,P_A(0)) & a_{e|\alpha} 
\end{array}
\right)
 \right],\label{JO1}
\end{align}
where
$c(\alpha,P_A(0)):=
(1-2P_A(0))\sqrt{a_{o|\alpha}a_{e|\alpha}}$.
In particular, when $P_A(0)=P_A(1)=1/2 $,
the off-diagonal part is zero.
Hence, we have
\begin{align}
H(
\frac{1}{2}|\alpha \rangle \langle \alpha  |
+\frac{1}{2}|-\alpha \rangle \langle -\alpha  |) 
=h(a_{o| \alpha}) \label{VOA}.
\end{align}

For the calculation of $I(AB;Y)$,
we need to handle three states
$|0\rangle, |2\alpha \rangle,|-2\alpha \rangle$.
For this aim, instead of $|{even}_{\alpha}\rangle$, we introduce another state;
\begin{align}
|\overline{even}_{\alpha}\rangle&:=a_{\overline{e}|\alpha}^{-1/2} 
e^{-|\alpha|^2/2}\sum_{m=1}^{\infty} \frac{\alpha^{2m}}{\sqrt{(2m)!}}
|2m\rangle ,
\end{align}
where 
$a_{\overline{e}|\alpha}:=
e^{-|\alpha|^2} (\cosh |\alpha|^2-1)=
{e^{-|\alpha|^2}
\sum_{m=1}^{\infty} \frac{|\alpha|^{4m}}{(2m)!}}$.
The state $|\overline{even}_{\alpha}\rangle$ is orthogonal to 
$|0\rangle$ and $|{odd}_{\alpha}\rangle$.
The coherent state $|2\alpha\rangle$ is decomposed in another way as
\begin{align}
|\pm 2 \alpha\rangle
=
e^{-2|\alpha|^2}|0\rangle
+\sqrt{a_{\overline{e}|2\alpha}} |\overline{even}_{2\alpha}\rangle
\pm \sqrt{a_{o|2\alpha}} |odd_{2\alpha}\rangle,
\end{align}

Based on the basis $\{|0\rangle, |odd_{2\alpha}\rangle,
|\overline{even}_{2\alpha}\rangle \}$,
we find that
\begin{align}
&I(AB;Y)\nonumber\\
=&H[
P_A(0)P_B(0)|2\alpha \rangle\langle 2\alpha |
+P_A(1)P_B(1)|-2\alpha \rangle\langle -2\alpha | \nonumber\\
&+(P_A(0)P_B(1)+P_A(1)P_B(0))|0 \rangle\langle 0 |] \nonumber\\
=&
H\left[
 \left(
\begin{array}{ccc}
c_0& c_2 & c_3\\
c_2&p_{AB} a_{o| 2\alpha} & c_1 \\
c_3&c_1 & p_{AB} a_{\overline{e}|2\alpha} 
\end{array}
\right)
 \right],
\end{align}
where
\begin{align}
p_{AB}&:=P_A(0)P_B(0)+P_A(1)P_B(1) \nonumber\\
c_{0}&:=(P_A(0)P_B(1)+P_A(1)P_B(0))+
p_{AB}e^{-4|\alpha|^2} \nonumber\\
c_{1}&:=(P_A(0)P_B(0)-P_A(1)P_B(1))
\sqrt{a_{\overline{e}|2\alpha} a_{o| 2\alpha}} \nonumber\\
c_{2}&:=(P_A(0)P_B(0)-P_A(1)P_B(1))
\sqrt{ a_{o| 2\alpha}} e^{-2|\alpha|^2}\nonumber\\
c_{3}&:=(P_A(0)P_B(0) + P_A(1)P_B(1))
\sqrt{a_{\overline{e}|2\alpha} } e^{-2|\alpha|^2}.\nonumber
\end{align}
Therefore, 
$I({\cal W}_\alpha)$ is given as the following maximum;
\begin{align}
I({\cal W}_\alpha)=
\max_{P_A,P_B}
H\left[
 \left(
\begin{array}{ccc}
c_0& c_2 & c_3\\
c_2&p_{AB} a_{o| 2\alpha} & c_1 \\
c_3&c_1 & p_{AB} a_{\overline{e}|2\alpha} 
\end{array}
\right)
 \right].
 \label{Expression_Achievable_rates_SD}
 \end{align}

Since $W_{0,1}$ and $W_{1,0}$ are the same state,
the mutual information $I(AB;Y)$ is upper bounded by $\log 3$.
The three states
$W_{0,1}=W_{1,0}$, $W_{0,0}$, $W_{1,1}$ can be distinguished when $\alpha$ is sufficiently large.
Under this limit, 
the maximum of the mutual information $I(AB;Y)$ 
converges to 
\begin{align}
I_{\max,\infty}:=\max_{P_A,P_B} \eta \big( P_A(0)P_B(1)+P_A(1)P_B(0) \big)+
\eta  \big(P_A(0)P_B(0) \big)+\eta \big(P_A(1)P_B(1) \big),
\end{align}
where $\eta(t):= -t \log t$.
That is, under this limit, the transmission rate of SD codes converges to $\frac{I_{\max,\infty}}{2}$.

\subsection{Rates for COMP codes}
In the setting with COMP codes, the lower bound of $C({\cal W}_\alpha)$ given in Theorem \ref{T1} is calculated as
\begin{align}
&{C}({\cal W}_\alpha)
\ge \underline{{C}}({\cal W}_\alpha) =
I(A\oplus B; Y) \nonumber \\
=& 
H\left[
 \left(
\begin{array}{ccc}
\tilde{c}_0& 0 & \sqrt{a_{\overline{e}|2\alpha} } e^{-2|\alpha|^2}/2 \\
0& a_{o| 2\alpha}/2 & 0\\
\sqrt{a_{\overline{e}|2\alpha} } e^{-2|\alpha|^2}/2
&0 & a_{\overline{e}|2\alpha}/2 
\end{array}
\right)
 \right]
-\frac{1}{2}
H\left[
 \left(
\begin{array}{ccc}
e^{-4|\alpha|^2}& 0 & \sqrt{a_{\overline{e}|2\alpha} } e^{-2|\alpha|^2}\\
0& a_{o| 2\alpha}& 0\\
\sqrt{a_{\overline{e}|2\alpha} } e^{-2|\alpha|^2} &0 & a_{\overline{e}|2\alpha}
\end{array}
\right)
 \right] 
  \nonumber\\
=& 
H\left[
 \left(
\begin{array}{ccc}
\tilde{c}_0& \sqrt{a_{\overline{e}|2\alpha} } e^{-2|\alpha|^2}/2& 0  \\
\sqrt{a_{\overline{e}|2\alpha} } e^{-2|\alpha|^2}/2
& a_{\overline{e}|2\alpha}/2&0  \\
0& 0& a_{o| 2\alpha}/2 
\end{array}
\right)
 \right]
-\frac{1}{2} h(a_{o| 2\alpha})   \label{eq:rate_COMP}\\
\ge & 
H\left[
 \left(
\begin{array}{ccc}
\tilde{c}_0& 0 & 0\\
0& a_{o| 2\alpha}/2 & 0\\
0&0 & a_{\overline{e}|2\alpha}/2 
\end{array}
\right)
 \right]
-\frac{1}{2}
H\left[
 \left(
\begin{array}{ccc}
e^{-4|\alpha|^2}& 0 & 0\\
0& a_{o| 2\alpha}& 0\\
0&0 & a_{\overline{e}|2\alpha}
\end{array}
\right)
 \right] \nonumber \\
=&h (\tilde{c}_{0})-\frac{1}{2}h(e^{-4|\alpha|^2}).
\label{eq:comprateBPSK}
\end{align}
Under the above relation, we consider the uniform distributions for $A$ and $B$.
Hence, \eqref{NJ2} follows from \eqref{eq:rate_COMP} while
the exact value of the optimal transmission rate ${C}({\cal W})$ is not known.

Since the three states
$W_{0,1}=W_{1,0}$, $W_{0,0}$, $W_{1,1}$ can be distinguished when $\alpha$ is sufficiently large.
Under this limit, 
the maximum of the mutual information $I(A \oplus B;Y)$ 
converges to $\log 2$.
Since $\log 2$ is upper bound of ${C}({\cal W})$,
we can conclude that 
${C}({\cal W})$ converges to $\log 2$ under this limit.

Further, the rate $h (\tilde{c}_{0})-\frac{1}{2}h(e^{-4|\alpha|^2})$ can be achieved without use of collective measurement as follows.
That is, it can be attained even with the following method.
The receiver applies the on-off measurement to each signal state, and obtains the outcome $X$, where 
the ``on'' corresponds to $1$ and the ``off'' does to $0$.
The obtained channel is given as
\begin{align}
P_{X|A\oplus B}(0|0)&= 1, \quad P_{X|A\oplus B}(1|0)= 0 \\
P_{X|A\oplus B}(0|1)&= e^{-4|\alpha|^2}, \quad 
P_{X|A\oplus B}(1|1)= 1-e^{-4|\alpha|^2}
\end{align}
when $P_A$ and $P_B$ are the uniform distribution.
Since the mutual information $I(A\oplus B;X)$ equals $h (\tilde{c}_{0})-\frac{1}{2}h(e^{-4|\alpha|^2})$, 
the rate $h (\tilde{c}_{0})-\frac{1}{2}h(e^{-4|\alpha|^2})$ can be attained by 
the combination of a classical encoder, a classical decoder, and the above on-off measurement.
That is, this rate can be achieved with a feasible measurement without quantum memory, and hence, we obtain \eqref{JO3}.

Next, we discuss the upper bound $\overline{{C}}({\cal W}_\alpha)$. We have
\begin{align}
&\max_{P_B, a\in \bF_2}I(B; Y|A=a)_{P_B}
=\max_{P_B, a\in \bF_2}
H\Big(\sum_{b\in \in \bF_2}P_B(b) |((-1)^a+(-1)^b)\alpha \rangle \langle ((-1)^a+(-1)^b)\alpha|\Big) \nonumber\\
=&\max_{P_B}
H\Big(\sum_{b\in \in \bF_2}P_B(b) |(-1)^b\alpha \rangle \langle (-1)^b\alpha|\Big) \nonumber\\
\stackrel{(a)}{=}& \max_{P_B}
H\left[
 \left(
\begin{array}{cc}
a_{o| \alpha} & c(\alpha,P_B(0)) \\
c(\alpha,P_B(0)) & a_{e|\alpha} 
\end{array}
\right)
 \right] 
=
H\left[
 \left(
\begin{array}{cc}
a_{o| \alpha} & 0 \\
0 & a_{e|\alpha} 
\end{array}
\right)
 \right] =h(a_{o|\alpha}),
\end{align}
where $(a)$ follows from \eqref{JO1}.
Notice that the quantity in the second line equals the single-sender Holevo quantity.
In the same way, we have
\begin{align}
\max_{P_A, b\in \bF_2}I(A; Y|B=b)_{P_A}
=H\left[
 \left(
\begin{array}{cc}
a_{o| \alpha} & 0 \\
0 & a_{e|\alpha} 
\end{array}
\right)
 \right] .
\end{align}
Thus, \eqref{NJ1} holds.

\subsection{Case of conventional homodyne detection}
As a more feasible detection, we focus on conventional homodyne detection with BPSK 
that has an outcome in the continuous set $\mathbb{R}$. 
It corresponds to measuring one quadrature of the annihilation field operator in \eqref{RecOP} with the outcome being the classical random variable defined as
\begin{align}
	Y =((-1)^A+(-1)^B)|\alpha| +Z,
\end{align}
where $A$ and $B$ are input variables of Senders A and B,
and $Z$ is the real-valued Gaussian variable with mean zero and variance $1/4$. 
In this case, the achievable rate of the COMP code is \cite{Ullah,Takabe,Takabe2}, \cite[(17)]{HV}
\begin{align}
I(A\oplus B; Y) 
&=\tilde{H}({Y}) - \tilde{H}({Y} | A \oplus B) \nonumber \\
&=h_1-\frac{1}{2}h_2-\frac{1}{4}.
\label{eq:Expression_Achievable_rates_COMP_hom}
\end{align}
where $\tilde{H}$ is the differential entropy, and 
$h_1$ and $h_2$ are defined by using $u(x):= -x \log x$ as 
\begin{align}
h_1&:=\int_{-\infty}^\infty\frac{\sqrt{2}}{\sqrt{\pi}}
u\Big(
\frac{1}{4} e^{-2(y-2 |\alpha|)^2} 
+\frac{1}{4} e^{-2(y+2 |\alpha|)^2} 
+\frac{1}{2} e^{-2y^2}
\Big) dy \nonumber \\
h_2&:=\int_{-\infty}^\infty\frac{\sqrt{2}}{\sqrt{\pi}}
u\Big(
\frac{1}{2} e^{-2(y-2|\alpha|)^2} 
+\frac{1}{2} e^{-2(y+2|\alpha|)^2} 
\Big) dy.
\end{align}

 \subsection*{Acknowledgments.}
MH was supported in part by 
Guangdong Provincial Key Laboratory (Grant No. 2019B121203002).

\appendices

\section{Application to butterfly network}
\begin{figure}[tbh]
    \centering
    \includegraphics[scale=0.5]{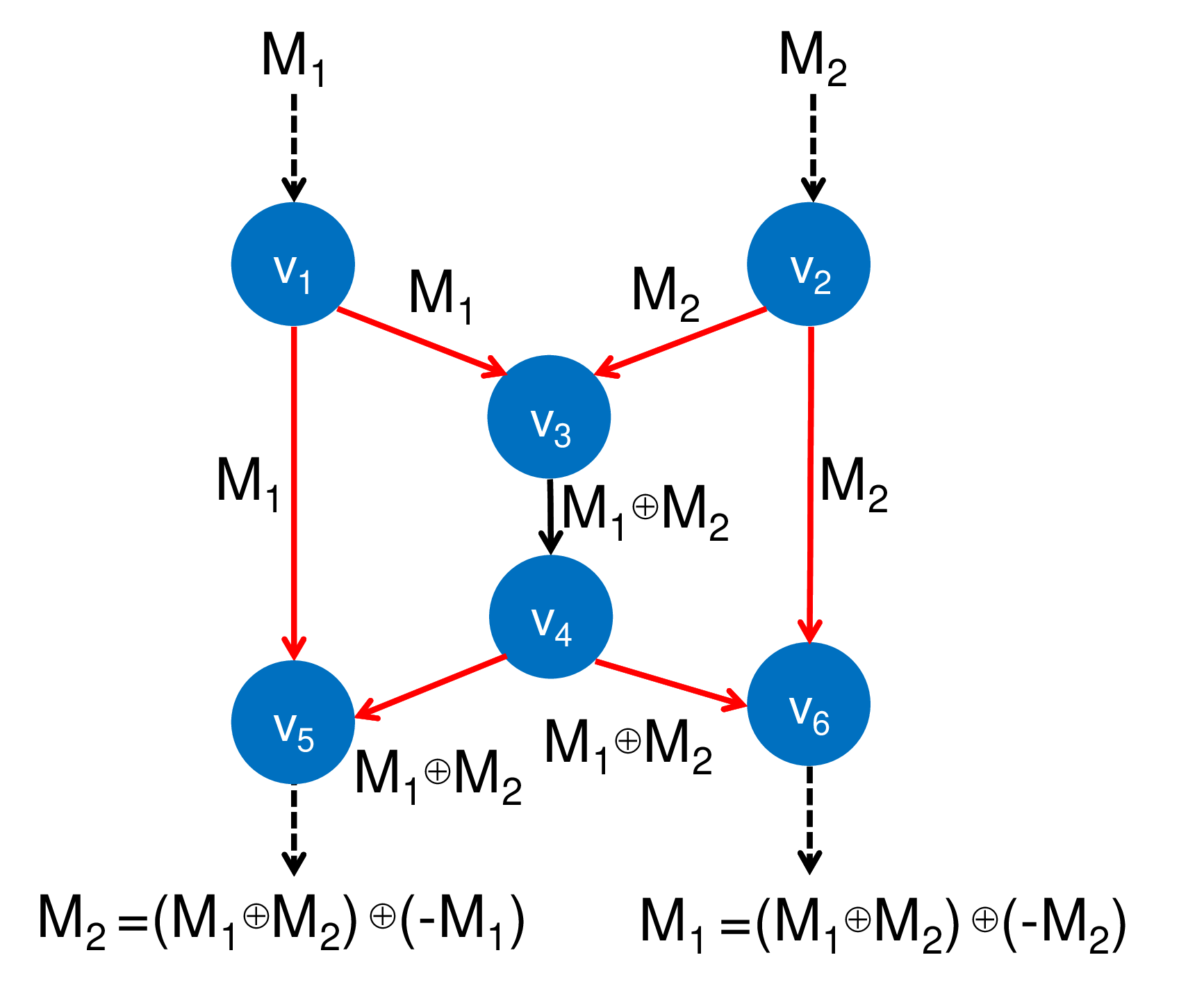}
    \caption{Another example where COMP codes offer an advantage, the butterfly network.
    The node $v_6$ ($v_5$) intends to receive the message from the node $v_1$ ($v_2$).
    In the optimal method, the nodes $v_3$, $v_5$, and $v_6$ recovers only the modulo sum of the two informations.
    A red arrow shows a part of cq-MAC, in which, our COMP code can be applied.
    }
    \label{butterfly}
\end{figure}

 As another example of the network topology, 
we can consider the butterfly network, shown in Fig. \ref{butterfly}.
While the papers \cite{Hayashi2007,PhysRevA.76.040301} address it, they assume noiseless channels.
In this network, the node $v_1$ ($v_2$) intends to the message to the node
$v_6$ ($v_5$).
The optimal method is presented in Fig. \ref{butterfly}.
To save the time, it is natural that the node $v_3$ receives the information from 
the two node $v_1$ and $v_2$ simultaneously, which implies that
the channel to the node $v_3$ is a cq-MAC.
Since the node $v_3$ needs only the modulo sum of these information,
we can apply our COMP code.
This method can be applied to 
the channels to the nodes $v_5$ and $v_6$ in the same way.
That is, our COMP code can be applied three times in the butterfly network.

\section{Analysis on affine codes}
\subsection{Optimal transmission rate with single sender}\label{LPT}
As a preparation for our proof of Theorem 1 of the main body, we 
analyze affine codes over a classical-quantum channel $W=\{W_x\}_{x \in {\cal X}}$ from the classical alphabet ${\cal X}$ to the quantum system ${\cal H}_Y$, 
where ${\cal X}$ is a finite field $\bF_q$ with prime power $q$.
Here, $W_x$  for $x \in {\cal X}$ is a density matrix on ${\cal H}_Y$.
The $n$-fold extension $W^{(n)}$ is defined so that
$W^{(n)}_{x^n}$ is the product state
$W_{x_1}\otimes \cdots \otimes W_{x_n}$ for $x^n=(x_1, \ldots, x_n)
\in {\cal X}^n$.
Let $I(X;Y)_{\rm uni}$ 
be the mutual information between $X$ and $Y$ when $X$ is subject to the uniform distribution.

In this section, for low-complexity construction of our codes,
our encoder is limited to a code constructed by affine maps.
A map $f$ onto a vector space
is called affine when it is written as 
$f(x)=A (x)\oplus b$ with a linear map $A$ and a constant vector $b$. 

\begin{theorem}\label{Pro12}
The mutual information $I(X;Y)_{\rm uni}$
equals the maximum transmission rate of 
an affine code with asymptotic zero error probability.
\end{theorem}

To show the achievability part of Theorem \ref{Pro12},
we randomly generate linear code with coding length $\ell$ as follows.
We randomly choose a linear function $F$ from $\bF_q^{\ell}$ to $\bF_q^n$
such that 
\begin{align}
{\rm Pr} \{ x \in {\rm Im} F \} \le q^{\ell-n} \hbox{ for } x (\neq 0)\in \bF_q^n,
\label{H40}
\end{align}
where ${\rm Im} F$ expresses the image of the function of $F$. 
Such a construction of ensemble of linear function $F$ is given by using Toeplitz matrix in \cite{TH13}.
We also randomly choose the shift vector $E\in \bF_p^{n}$ subject to the uniform distribution as an variable independent of $F$.
In the following, using ${F}$ and $E$,
we define our affine code $G$ as $G(m):= F(m)+E$.
Then, we define our decoder for the code $G$ as follows.
For $x^n \in {|cal X}^n$, we define the projection
\begin{align}
\Pi_{x^n}:= \{ W^{(n)}_{x^n} \ge q^{\ell}  W_{mix}^{\otimes n}\},
\end{align}
where $W_{mix}:= \sum_{x \in \bF_q}\frac{1}{q} W_x$.
Then, we define the decoder $\{D_{x^n,G}\}_{ x^n \in \im G  }$
\begin{align}
D_{x^n,G}:= \Big(\sum_{ {x^n}' \in \im G }\Pi_{{x^n}'}\Big)^{-1/2}
\Pi_{x^n}\Big(\sum_{ {x^n}' \in \im G }\Pi_{{x^n}'}\Big)^{-1/2}.
\Label{PT}
\end{align}
We denote the decoding error probability of the code $G$ with the above decoder
by $\epsilon(G)$.
\begin{lemma}\label{Pro1}
When we generated our affine code $G$ as the above,
the average of the decoding error probability $\epsilon(G)$ is upper bounded as
\begin{align}
\mathbb{E}_G \epsilon(G)\le 6
q^{s \ell}  e^{n E(s)}
\end{align}
with $s \in (0,1)$
where 
$E(s):=\log \Big(\sum_{x \in \bF_q}q^{-1}\Tr W_{x}^{1-s}W_{mix}^s\Big)$.
\end{lemma}
This proposition shows the achievability part of Proposition \ref{Pro12}
because $E(s)\cong s I(X;Y)_{\rm uni}$
when $s$ is close to $0$.

Conversely, 
the impossibility part of Theorem \ref{Pro12}
can be shown as follows.
Let $g$ be an affine code with block length $n$.
Assume $M$ is the message subject to the uniform distribution and 
$X^n=(X_1, \ldots, X_n)$ is given as $g(M)$.
The mutual information $I(X^n;Y^n)$ is evaluated by
\begin{align}
&I(X^n;Y^n)=\sum_{i=1}^n I(X^n;Y_i|Y_1,\ldots,Y_{i-1}) 
\le \sum_{i=1}^n I(X^n,Y_1,\ldots,Y_{i-1};Y_i)
= \sum_{i=1}^n I(X_i;Y_i)\label{HK1}
\end{align}
because of the Markov chain $(X^n,Y_1,\ldots,Y_{i-1})-X_i-Y_i$.
Since $g$ is an affine map,
$X_i$ is subject to the uniform distribution unless it is a constant.
Hence, $I(X_i;Y_i)$ is upper bounded by $I(X;Y)_{\rm uni}$.
Combining Fano inequality, we can show the above fact.

\subsection{Proof of Lemma \ref{Pro1}}
Lemma \ref{Pro1} can be shown as follows.
Assume the following condition for our code $G$;
\begin{description}
\item[C-$x^n$] The element $x^n \in {\cal X}^n$ is chosen as a code word of the code $G$.
\end{description}
Then, the condition \eqref{H40} yields the following inequality for ${x^n}'(\neq x^n) \in \bF_q^n$;
\begin{align}
{\rm Pr}_{C-x^n} \{ {x^n}'\in {\rm Im} G \} \le q^{\ell-n} ,
\label{H40}
\end{align}
where ${\rm Pr}_{C-x^n}$ expresses the conditional probability with the condition C-$x^n$.

By using Hayashi-Nagaoka inequality \cite{HN},  the decoding error probability 
for $x^n \in {\cal X}^n$ is evaluated as 
\begin{align}
\Tr W^{(n)}_{x^n} (I-D_{x^n,G})
\le 2 \Tr W^{(n)}_{x^n} (I-\Pi_{x^n})
+ 4 \Tr W^{(n)}_{x^n} \Big(\sum_{ {x^n}' (\neq x^n)\in \im G }\Pi_{{x^n}'}\Big).
\end{align}
Then, 
\begin{align}
& \mathbb{E}_G \Big[
\sum_{x^n \in \im G}q^{-\ell}
\Tr W^{(n)}_{x^n} (I-\Pi_{x^n})\Big] \nonumber \\
\le &
\mathbb{E}_G \Big[
\sum_{x^n \in \im G}q^{-\ell}
\Tr (W^{(n)}_{x^n})^{1-s}
q^{s\ell}  (W_{mix}^{\otimes n})^s\nonumber \\
=&
q^{s\ell}
\Big(\sum_{x \in \bF_q}q^{-1}\Tr W_{x}^{1-s}W_{mix}^s\Big)^n
=q^{s \ell}  e^{n E(s)}.
\end{align}
Hence, we have
\begin{align}
&\mathbb{E}_G \Big[
\sum_{x^n \in \im G}q^{-\ell}
\Tr W^{(n)}_{x^n} \Big(\sum_{ {x^n}' (\neq x^n)\in \im G }\Pi_{{x^n}'}\Big)
\Big]\nonumber \\
=&
\mathbb{E}_G \Big[
q^{-\ell}
\sum_{ {x^n}' \in \im G }
\Tr \Pi_{{x^n}'}
\Big(
\sum_{x^n (\neq {x^n}') \in \im G}
W^{(n)}_{x^n} 
\Big)
\Big]\nonumber \\
=&
\mathbb{E}_G \Big[
q^{-\ell}
\sum_{ {x^n}' \in \im G }
\Tr \Pi_{{x^n}'}
\mathbb{E}_{G|C-x^n} 
\Big(
\sum_{x^n (\neq {x^n}') \in \im G}
W^{(n)}_{x^n} 
\Big)
\Big]\nonumber \\
=&
\mathbb{E}_G \Big[
q^{-\ell}
\sum_{ {x^n}' \in \im G }
\Tr \Pi_{{x^n}'}
\Big(
\sum_{x^n (\neq {x^n}') \in \bF_q^n}
{\rm Pr}_{C-x^n} \{ {x^n}'\in {\rm Im} G \}
W^{(n)}_{x^n} 
\Big)
\Big]\nonumber \\
\le &
\mathbb{E}_G \Big[
q^{-\ell}
\sum_{ {x^n}' \in \im G }
\Tr \Pi_{{x^n}'}
\Big(
\sum_{x^n (\neq {x^n}') \in \bF_q^n}
q^{\ell -n}
W^{(n)}_{x^n} 
\Big)
\Big]\nonumber \\
\le &
\mathbb{E}_G \Big[
q^{-n}
\sum_{ {x^n}' \in \im G }
\Tr \Pi_{{x^n}'}
\Big(
\sum_{x^n \in \bF_q^n}
W^{(n)}_{x^n} 
\Big)
\Big]\nonumber \\
= &
\mathbb{E}_G \Big[
q^{-n}
\sum_{ {x^n}' \in \im G }
\Tr \Pi_{{x^n}'}
q^n W_{mix}^{\otimes n}
\Big]\nonumber \\
\le &
\mathbb{E}_G \Big[
\sum_{ {x^n}' \in \im G }
\Tr (W^{(n)}_{{x^n}'})^{1-s}
q^{(1-s)\ell}  (W_{mix}^{\otimes n})^s
\Big]\nonumber \\
= &
q^{\ell - n}
\sum_{ {x^n}' \in \bF_q^n }
\Tr (W^{(n)}_{ {x^n}'})^{1-s}
q^{(1-s)\ell}  (W_{mix}^{\otimes n})^s
\nonumber \\
=&
q^{s\ell}
\Big(\sum_{x \in \bF_q}q^{-1}\Tr W_{x}^{1-s}W_{mix}^s\Big)^n
=q^{s \ell}  e^{n E(s)}.
\end{align}
Therefore, the average of the decoding error probability is evaluated as 
\begin{align}
\mathbb{E}_G \Tr W^{(n)}_{x^n} (I-D_{x^n,G})
\le 6 q^{s \ell}  e^{n E(s)}.
\end{align}
we obtain Lemma \ref{Pro1}.

\section{Proof of Theorem 1}\label{PfT1}
\subsection{Lower bound}
First, 
generalizing the method of \cite{Ullah} for classical channels
 to classical-quantum MACs, we prove the lower bound (1) of the main body. That is,
we show that the mutual information rate
$ I(Y ; A \oplus B)$ can be achieved in the sense of computation-and-forward.
To show this theorem, we define the degraded channel $\{W_{x|d}\}_{x \in \bF_p}$
from $\bF_p$ to ${\cal H}_Y$ as
$W^d_{x}:=\sum_{x'\in \bF_p} p^{-1}W_{x',x\ominus x}$ for $x \in \bF_p$.
When $X$ is subject to the uniform distribution on $\bF_p$
and the output state  is given as $W_{X|d}$,
we have $ I(Y ; A \oplus B)_{\rm uni}=I(Y;X)$.

We randomly generate our code with coding length $\ell$. 
We randomly choose a linear function $F$ as \eqref{H40} in the same way as Section \ref{LPT}.
We extend the function $F$ to an invertible linear function $\bar{F}$ on $\bF_p^n$
such that the restriction of $\bar{F}$ to the first $\ell$ component equals the function $F$.
We also randomly choose the shift vectors 
$E\in \bF_p^{n}$ and 
$E'\in \bF_p^{n-\ell}$ independently subject to the uniform distribution as other variables.
In the following, using $\bar{F}$, $E$, and $E'$,
we construct our code as follows.
That is, our code depends on $\bar{F}$, $E$, and $E'$.
Also, we define the affine code $G$ as $G(m):=F(m)\oplus E $.
 
The sender A encodes the message $M_A^n$ to 
$\bar{F}(M_A^n,E')$,
and  
the sender B encodes the message $M_B^n$ to 
$\bar{F}(M_B^n,-E')+E$.
In the following, we assume that the receiver
makes the decoder dependently only of $\bar{F}$, ${F}$ and $E$.
Hence, $\bar{F}$, $F$, $E$ are fixed to $\bar{f}$, $f$, and $e$.
Also, $g$ express the affine code defined as $g(m):=f(m)\oplus e $.

When $M_A^n\oplus M_B^n=m$, the output state is
\begin{align}
& \mathbb{E}_{M_A^n,E' }W^{(n)}_{\bar{f}(M_A^n,E'),\bar{f}(m\ominus M_A^n,-E')\oplus e }
=
\mathbb{E}_{M_A^n,E' }W^{(n)}_{\bar{f}(M_A^n,E'),\bar{f}(m)\oplus e\ominus \bar{f}(M_A^n,E') } \nonumber \\
= &
q^{-n}\sum_{x^n \in \bF_p^n } W^{(n)}_{x^n,\bar{f}(m)\oplus e\ominus x^n }
=W^{d,(n)}_{\bar{f}(m)\oplus e}
=W^{d,(n)}_{{f}(m)\oplus e}.
\end{align}
Hence, the above state is the output state with the following case;
The encoder is the affine code $g$ and the channel is the degraded channel $W^d$.
Given the affine code $g$, we choose the decoder based on \eqref{PT}
when the channel is the degraded channel $W^d$.

Now, we consider the case when $F$ and $E$ are randomly chosen in the above way.
This random choice of $G$ is the same as Lemma \ref{Pro1}.
Hence, Lemma \ref{Pro1} guarantees that 
the rate $ I(Y ; A \oplus B)_{\rm uni}=I(Y;X)$ is achievable, which implies 
(1) of the main body.

\subsection{Upper bound}
Next, we prove the upper bound (2) of the main body.
For this aim, we focus on a sequence of codes $\{\Psi_{n}\}$
with a transmission rate pair $R$, where
the pair of encoders $(\phi_{A,n},\phi_{B,n})$ of $\Psi_{n}$ satisfies the following;
$\phi_{A,n}(\phi_{B,n})$ maps $M_{A,n}(M_{B,n})$ to $X_A^n=(X_{A,i})_{i=1}^n
(X_B^n=(X_{B,i})_{i=1}^n)$.
Here, $M_{A,n}$ and $M_{B,n}$ are independently subject to the uniform distribution.
Since
\begin{align}
& I(M_{A,n}\oplus M_{B,n};Y^n|M_{A,n} ) \nonumber \\
=& I(M_{A,n}\oplus M_{B,n};Y^n M_{A,n} )-
I(M_{A,n}\oplus M_{B,n};M_{A,n} ) \nonumber \\
=& I(M_{A,n}\oplus M_{B,n};Y^n M_{A,n} ) 
\ge  I(M_{A,n}\oplus M_{B,n};Y^n  ) ,
\end{align}
we find that
\begin{align}
& I(M_{A,n}\oplus M_{B,n};Y^n)
{\le}  I(I(M_{A,n}\oplus M_{B,n};Y^n|M_{A,n} )
=  I(M_{B,n};Y^n|M_{A,n} ) =  I(X_B^n;Y^n|X_A^n ) \nonumber \nonumber \\
=& \sum_{i=1}^n  I(X_B^n;Y_i| Y^{i-1}X_A^n)
= \sum_{i=1}^n  I(X_{B,i};Y_i| Y^{i-1}X_A^n)
\stackrel{(a)}{\le} \sum_{i=1}^n  I(X_{B,i};Y_i| X_A^n),
= \sum_{i=1}^n  I(X_{B,i};Y_i| X_{A,i}),
\end{align}
where $(a)$ follows from the Markov chain 
$Y^{i-1}-X_{B,i}-Y_i$ when $X_A^n$ is fixed.

Combining Fano's inequality, we can show that
\begin{align}
R &\le 
\max_{P_A , P_B}I(B; Y|A)_{P_A \times P_B}=\max_{P_B, a\in \bF_p}I(B; Y|A=a)_{P_B}.
\end{align}
In the same way, we can show
\begin{align}
R &\le \max_{P_A , P_B}I(A; Y|B)_{P_A \times P_B}
=\max_{P_A, b\in \bF_p}I(A; Y|B=b)_{P_A}.
\end{align}

\end{document}